\documentclass[twocolumn,10pt]{revtex4}  
\usepackage{shortcuts}
\usepackage{amsmath}
\usepackage{psfig}
\usepackage{psfrag}
\usepackage{graphics}
\usepackage{graphicx}

\begin{document}                

\title{Dielectric properties of a two dimensional binary sytem with ellipse inclusions}
\author{{\me}}
\email{enis.tuncer@elkraft.chalmers.se}
\affiliation{Chalmers University of Technology, SE-412\ 96 Gothenburg Sweden}

\begin{abstract}                
A two-dimensional binary composite system composed of ellipse inclusions and a host medium is considered. Dielectric permittivity of the system is calculated as a function of orientation angle, the volume fraction of the inclusions and their excentricity using the finite element method. It was observed that both the orientation of the inclusions in the field and their excentricity have significant effects on the dielectric permittivity. 
\end{abstract}
\maketitle

{B}{inary} systems composed of two media have been studied through the history due to their simplicity. Dielectric properties of such systems show different behaivour depending on the distribution and shape of the inclusions \cite{Sillars1937,Landauer1978}. Materials used in electrical insulation sytems are usually composed of several materials. Therefore, behavior of such systems in electrical fields are crucial for the insulation.  

In this paper, we determine the effective parameters of a two diemsional structure composed of a main phase and ellipse inclusions. We have assumed that the structure is periodic where the  inclusions sit on lattice points of a square array. With such an assumption the system can be subdivided into smaller units \ie square cells. This approach was first introduced by Rayleigh \cite{Rayleigh} and it has been used in numerical and analytical techniques \cite{Sareni1996,Tuncer1998b}. In this model, inclusions were arranged on rectangular lattice points. One can assume cylinderical inclusions in two-dimensions and pherical inclusions in three dimesions. Analytical solutions for two component and three component systems with different inclusion geometries (infintely long cylinders) were given in Refs. \cite{pier6,Emets98b,Steeman90}. Moreover, frequency dependent characteristics, lattice effects and random structures were presented in Refs. \cite{Tuncer1998b,Tuncer1999a}.

Mean field theory approaches and solutions for frequency dependent effective medium properties of a mixture with ellipsoidal inclusions were studied by Sillars \cite{Sillars1937} and Steeman and Maurer \cite{Steeman1992}. The former one assumed that the inclusions were much more conductive than the matrix material, and the later one introduced the conductivity of the matrix. In both cases they have assumed that the ellipsoidals were parallel or perpendicular to the applied electric field. Therefore, we concentrated on the orientation of the ellipses in the periodic structure. First,  all the ellipses were assumed to have the same orientation. Later, we change the orientation of the neighbor ellipse which made the system doubly periodic. We exclude the effects of ellipse concentration on effective parameters in the mixture.

\begin{figure}[pbt]
\begin{center}
\begin{tabular}{c}
\psfragscanon
\psfrag{ep}[][l]{$\varepsilon_m$}
\psfrag{ef}[][l]{$\varepsilon_i$}
\psfrag{sp}[][r]{$\sigma_m$}
\psfrag{sf}[][r]{$\sigma_i$}
\psfrag{1}[][r]{$1$}
\psfrag{2}[][r]{$2$}
\psfrag{3}[][r]{$3$}
\psfrag{4}[][r]{$4$}
\includegraphics[width=6.5cm,angle=-90]{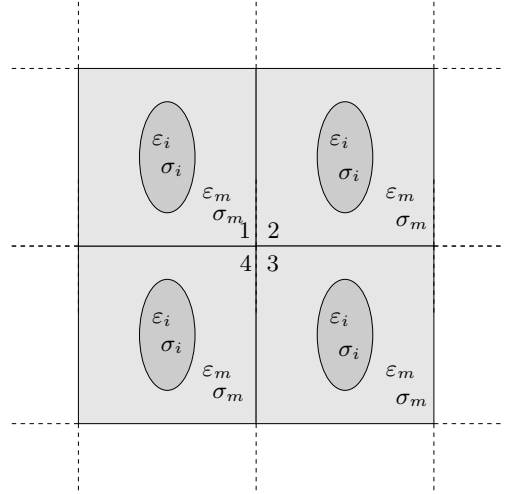} \\
\psfragscanoff
\end{tabular}
\end{center}
\caption{Unit cell.}
\label{unitcell}
\end{figure}

Let ellipses with dielectric permittivities $\varepsilon_i$ be located in an unbounded medium with $\varepsilon_m$. The ellipses are placed at the square lattice sites as presented in Fig. \ref{unitcell}. The solution of the Maxwell equations without any magnetic field presence, will assist to estimate the effective properties of the mixture. Assumptions of the ellipse in two dimensions and the ellipsoidal  in three dimensions can explain effective properties of composite structures better than considerations with spherical or cylinderical inclusions, since real composites have arbitrary shaped inclusions. Moreover, the interaction between ellipse disks will be different depending on the orientation of the neighboring ellipses. 

\begin{figure}[pbt]
\begin{center}
\psfragscanon
\psfrag{b}[][l]{$\beta$}
\psfrag{a}[][l]{$\alpha$}
\psfrag{B}[][l]{$a$}
\psfrag{A}[][l]{$b$}
\psfrag{L(x,y)}[][r]{$L(x,y)$}
\psfrag{x}[][r]{$x$}
\psfrag{y}[][r]{$y$}
\includegraphics[width=6.5cm,angle=-90]{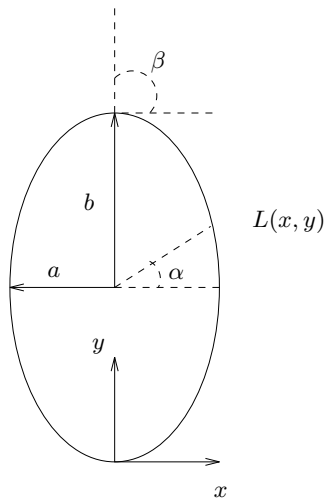} \\
\psfragscanoff
\end{center}
\caption{Geometrical parameters for ellipses.}
\label{elips}
\end{figure}

\begin{figure}
\begin{center}
\psfragscanon
\psfrag{x}[][l]{$a$}
\psfrag{y}[][l]{$b$}
\psfrag{q}[r][r]{$q$}
\psfrag{0.05}[][r]{}
\psfrag{0.15}[][r]{}
\psfrag{0.25}[][r]{}
\psfrag{0.35}[][r]{}
\psfrag{0.45}[][r]{}
\psfrag{0.1}[][c]{0.1}
\psfrag{0.2}[][c]{0.2}
\psfrag{0.3}[][c]{0.3}
\psfrag{0.4}[][c]{0.4}
\psfrag{0.5}[][c]{0.5}
\psfrag{0.11}[][cr]{0.1}
\psfrag{0.21}[][cr]{0.2}
\psfrag{0.31}[][cr]{0.3}
\psfrag{0.41}[][cr]{0.4}
\psfrag{0.51}[][cr]{0.5}
\psfrag{0.61}[][cr]{0.6}
\psfrag{0.71}[][cr]{0.7}
\psfrag{0}[][c]{0}
\includegraphics[width=6.5cm,angle=-90]{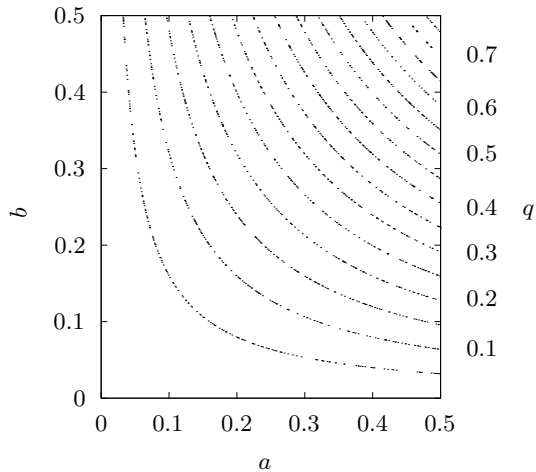} \\
\psfragscanoff
\end{center}
\caption{Equi-concentration lines.}
\label{equi-concentration}
\end{figure}

\begin{figure}[b]
\begin{center}
\psfragscanon
\psfrag{x}[][l]{$a$}
\psfrag{y}[][l]{$b$}
\psfrag{z}[r][r]{$e$}
\psfrag{0.1}[][c]{0.1}
\psfrag{0.2}[][c]{0.2}
\psfrag{0.3}[][c]{0.3}
\psfrag{0.4}[][c]{0.4}
\psfrag{0.6}[][c]{0.6}
\psfrag{0.8}[][c]{0.8}
\psfrag{1}[][c]{1}
\psfrag{0}[][c]{0}
\includegraphics[width=6.5cm,angle=-90]{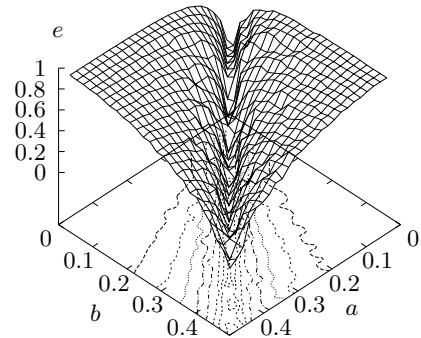} \\
\psfragscanoff
\end{center}
\caption{Eccentricity as a function of $a$ and $b$.}
\label{eccentricity}
\end{figure}

Unit ellipse in $xy$-coordinates is given by (Fig. \ref{elips})
\begin{equation}     
\frac{(x-x_0^2)}{a^2}+\frac{(y-y_0)^2}{b^2}=1
\end{equation}
where $x_0$ and $y_0$  are the center coordinates of the ellipse, and $a$ and $b$ are the maximum radii in $x$ and $y$ directions. One can also use polar-coordinates,
\begin{equation}
\begin{bmatrix}
x\\y
\end{bmatrix}
=
\begin{bmatrix}
a \cos\alpha\\ b \cos\alpha
\end{bmatrix}
+
\begin{bmatrix}
x_0 \\ y_0
\end{bmatrix}
\end{equation}
Moreover, if one includes the orientation of the ellipse, angle from the $x$-axis $\beta$, in the matrix medium, the $x$ and $y$ points will be multiplied by the transformation matrix,
\begin{equation}
\begin{bmatrix}
x\\y
\end{bmatrix}
=
\begin{bmatrix}
 \cos\beta & -\sin\beta \\ \sin\beta & \cos\beta \\
\end{bmatrix}
\begin{bmatrix}
a \cos\alpha\\ b \cos\alpha
\end{bmatrix}
+
\begin{bmatrix}
x_0 \\ y_0
\end{bmatrix}
\end{equation}
A point, $L(x,y)$ can be calculated in this way.
The shape factor, eccentricity ($e$), of the ellipse is
\begin{equation}     
e=\frac{\sqrt{a^2-b^2}}{a}
\end{equation}
In this equation, the boundaries of the unit cell as well as the area fraction  of the inclusion ellipse are the constraints for the values of $a$ and $b$. Since, the ellipse inclusions are hard disk-like structures, they are not allowed to overlap. This leads to $a<1/2$ and $b<1.2$ in the unit cell. The resulting  $a$ and $b$ values versus equi-concentration lines and the eccentricity are presented in Fig.\ \ref{equi-concentration} and Fig.\ \ref{eccentricity}. However, for orientation angles, $\beta$ other than $\beta\ne n\pi/2$, $n$ being an integer, the previous constraints will be different, which is not included in this paper. The area fraction, $q$, of ellipse in the unit square medium is
\begin{equation}     
q=\pi ab
\end{equation}     

\begin{figure}
  \psfragscanon
  \psfrag{LOG(OMEGA)}[c][c]{$\log(\omega)$}
  \psfrag{LOG(CHI)}[c][c]{$\log(\chi)$}
  \psfrag{90-90-a40-RE}[c][r]{$\textsf{I}$}
  \psfrag{90-90-a40-IM}[c][r]{$\textsf{II}$}
  \psfrag{90-90-a35-RE}[c][r]{$\textsf{III}$}
  \psfrag{90-90-a35-IM}[c][r]{$\textsf{IV}$}
  \psfrag{3-RE}[c][r]{$\textsf{V}$}
  \psfrag{3-IM}[c][r]{$\textsf{VI}$}
  \includegraphics[width=6.5cm,angle=-90]{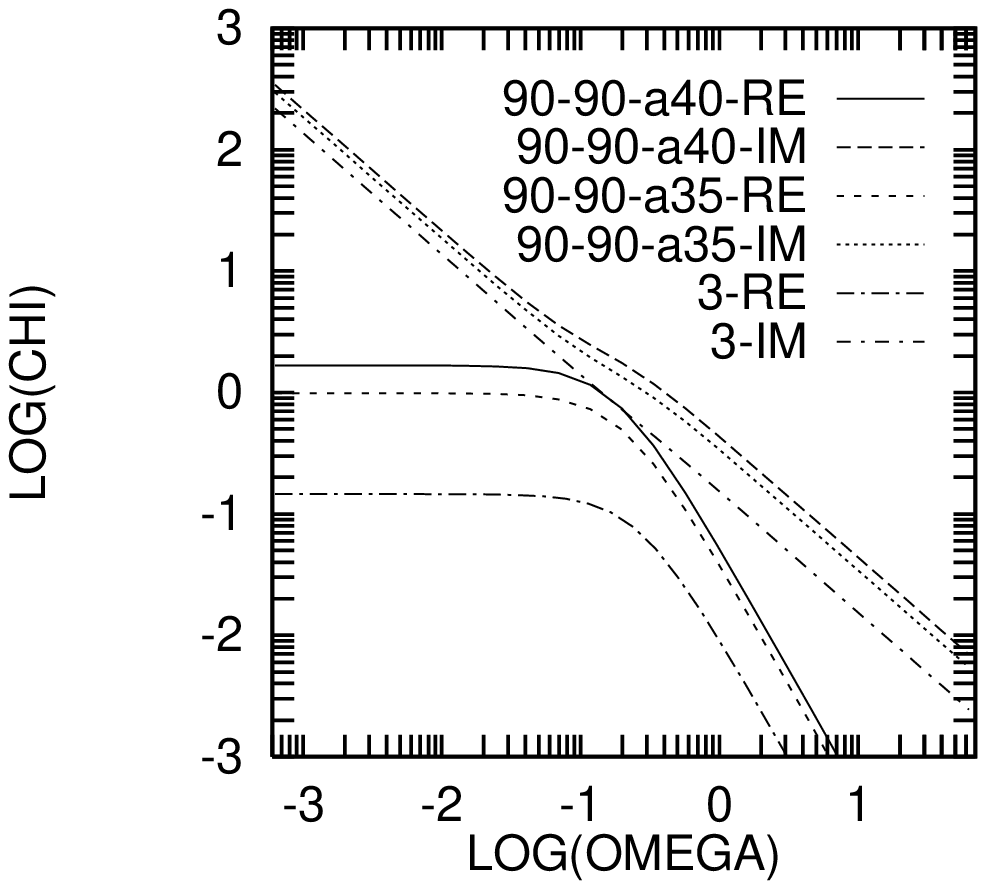}
  \psfragscanoff \\
  \psfragscanon
  \psfrag{LOG(OMEGA)}[c][c]{$\log(\omega)$}
  \psfrag{LOG(CHI)}[c][c]{$\log(\chi)$}
  \psfrag{0-0-a40-RE}[c][r]{$\textsf{I}$}
  \psfrag{0-0-a40-IM}[c][r]{$\textsf{II}$}
  \psfrag{0-0-a35-RE}[c][r]{$\textsf{III}$}
  \psfrag{0-0-a35-IM}[c][r]{$\textsf{IV}$}
  \psfrag{3-RE}[c][r]{$\textsf{V}$}
  \psfrag{3-IM}[c][r]{$\textsf{VI}$}
  \includegraphics[width=6.5cm,angle=-90]{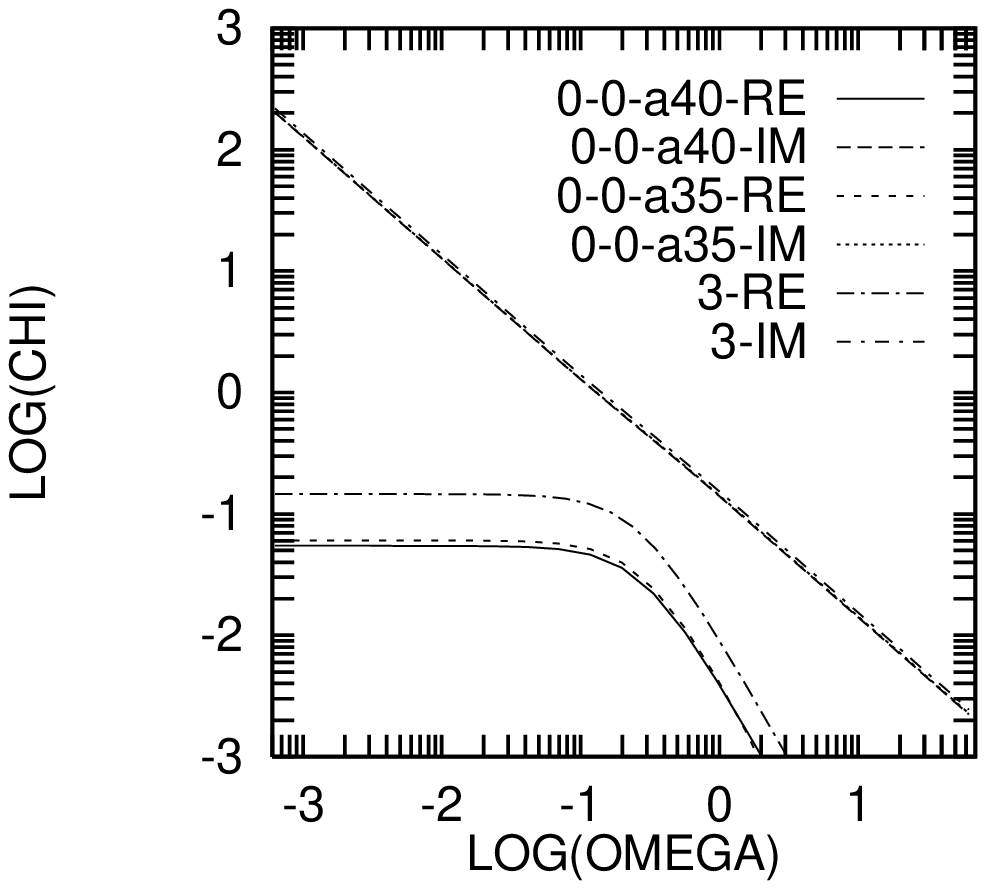}
  \psfragscanoff
  \caption{\label{fig:shape1}
Complex dielectric susceptibilities, $\chi^\ast$,  of a binary system with different ellipse orientations and eccentricities at $q=0.1$. The angle $\beta$ for neighboring ellipses is the same; (a) $\beta_{1,2}=\pi/2$ and (b) $\beta_{1,2}=0$. The labels $\sf{I}$ and $\sf{II}$ represent  $\chi'$ and $\chi''$-values for $e=0.980$. In a similar way, the labels $\sf{III}$ and $\sf{IV}$ represent $\chi'$ and $\chi''$-values for $e=0.965$. The labels $\sf{V}$ and $\sf{VI}$ represent $\chi'$ and $\chi''$-values for a circular inclusion, $e=0$.}
\end{figure}

To study the frequency dependent effective properties, for example, effective dielectric permittivity and conductivity, of the mixture, finite element method (FEM) numerical technique was used in two dimensions. The complex dielectric constant, $\varepsilon^\ast$ carries information about the intrinsic electrical properties \ie dc conductivity, $\sigma$ and polarization, $\varepsilon$ of the medium. Excluding the dipolar relaxations in the material, the complex dielectric constant becomes
\begin{align}
\begin{split}
\varepsilon^\ast = \varepsilon_s -i \frac{\sigma}{\varepsilon_0 \omega}\\
\end{split} 
\end{align}
When two materials form an interface that satisfy the condition $\varepsilon_1/\sigma_1\ne\varepsilon_2/\sigma_2$, a relaxation process due to the interfacial (Maxwell-Wagner-Sillars) polarization is observed. Since we have excluded the other relaxation processes coming from the materials, the dielectric permittivity of the mixture will be swithed after some frequency. It will be the same for the conductivity.  

\begin{figure}[t]
      \psfragscanon
      \psfrag{LOG(OMEGA)}[c][c]{$\log(\omega)$}
      \psfrag{LOG(CHI)}[c][c]{$\log(\chi)$}
      \psfrag{0-0-a40-RE}[c][r]{$\textsf{III}$}
      \psfrag{0-0-a40-IM}[c][r]{$\textsf{IV}$}
      \psfrag{90-90-a40-RE}[c][r]{$\textsf{I}$}
      \psfrag{90-90-a40-IM}[c][r]{$\textsf{II}$}
      \psfrag{3-RE}[c][r]{$\textsf{V}$}
      \psfrag{3-IM}[c][r]{$\textsf{VI}$}
      {\includegraphics[width=6.5cm,angle=-90]{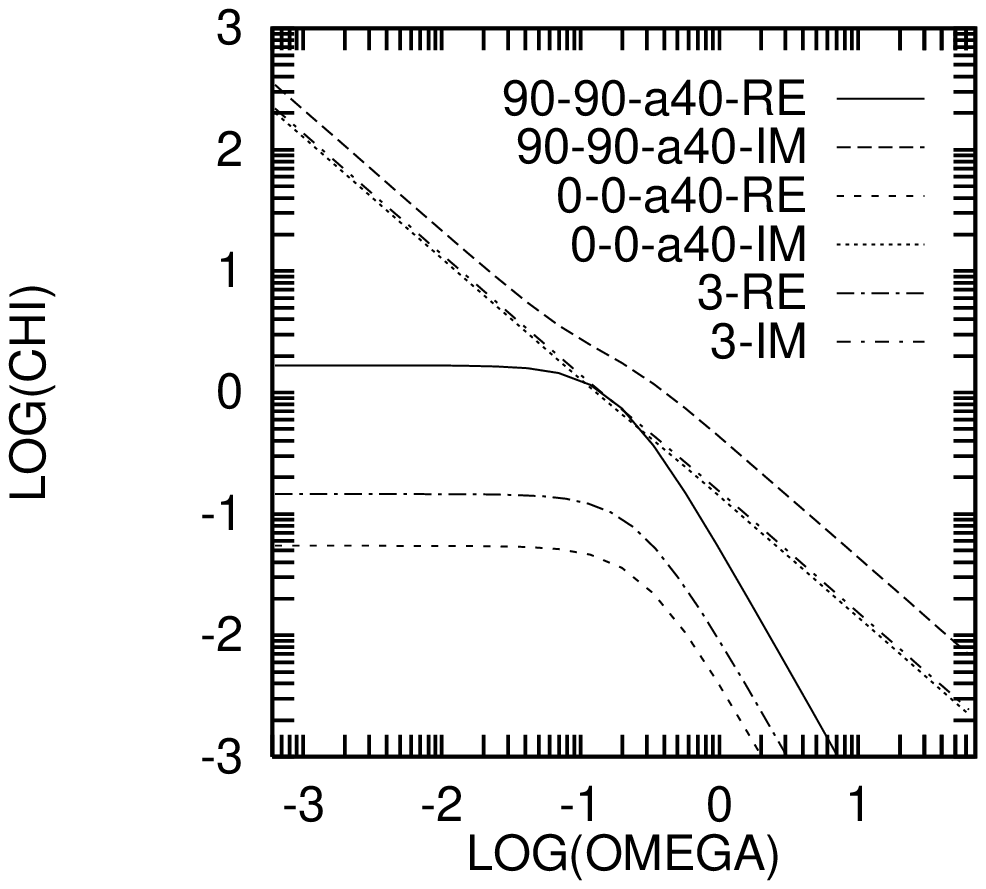}}
      \psfragscanoff \\
      \psfragscanon
      \psfrag{LOG(OMEGA)}[c][c]{$\log(\omega)$}
      \psfrag{LOG(CHI)}[c][c]{$\log(\chi)$}
      \psfrag{30-30-a40-RE}[c][r]{$\textsf{III}$}
      \psfrag{30-30-a40-IM}[c][r]{$\textsf{IV}$}
      \psfrag{60-60-a40-RE}[c][r]{$\textsf{I}$}
      \psfrag{60-60-a40-IM}[c][r]{$\textsf{II}$}
      \psfrag{3-RE}[c][r]{$\textsf{V}$}
      \psfrag{3-IM}[c][r]{$\textsf{VI}$}
      {\includegraphics[width=6.5cm,angle=-90]{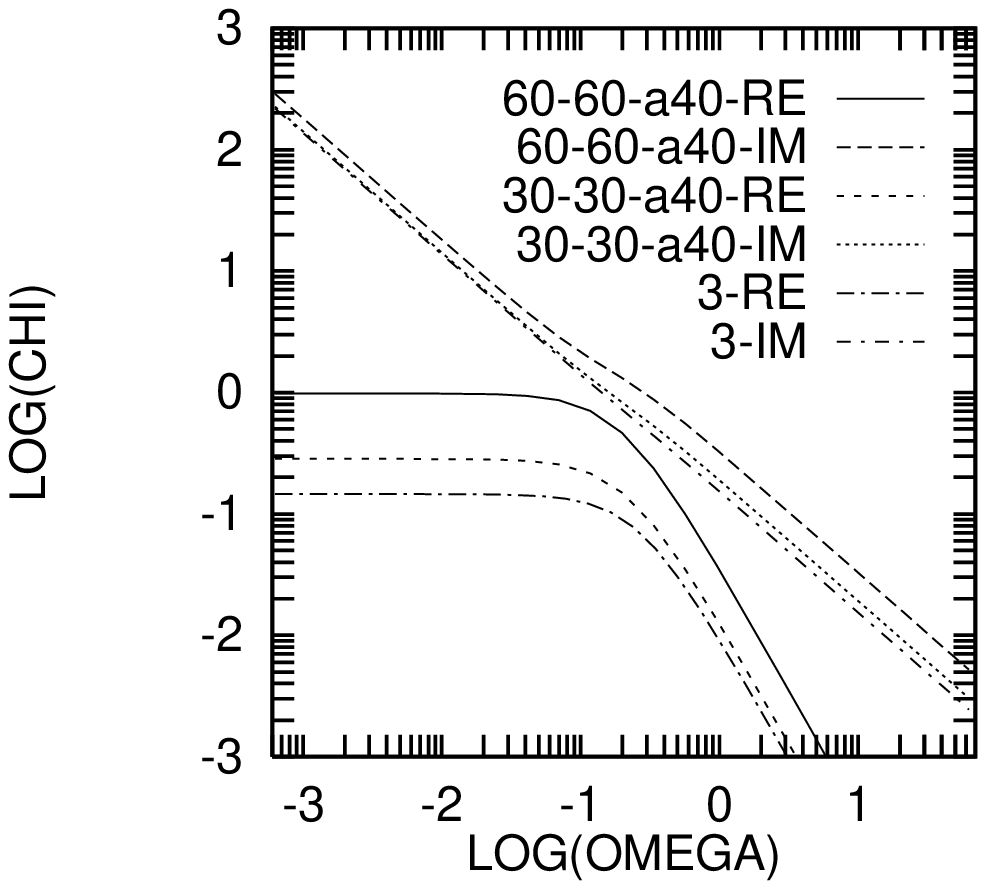}}
      \psfragscanoff
    \caption{\label{fig:shape2}Complex dielectric susceptibilities, $\chi^\ast$, of a binary system with different inclusion orientations at $q=0.1$ when $e=0.980$. The angle $\beta$ for neighboring ellipses is the same (a) $\beta_1=\beta_2=\pi/2$ (labels $\sf{I}$ and $\sf{II}$) and $\beta_1=\beta_2=0$ (labels $\sf{III}$ and $\sf{IV}$), and (b) $\beta_1=\beta_2=\pi/3$ (labels $\sf{III}$ and $\sf{IV}$) and $\beta_1=\beta_2=\pi/6$ (labels $\sf{III}$ and $\sf{IV}$). The labels $\sf{V}$ and $\sf{VI}$ are for a circular inclusion, $e=0$. The odd and even labels represent $\chi'$ and $\chi''$-values, respectively.}
\end{figure}
The calculation of the effective properties of the mixture from the known material properties of the constituents are electromagnetic problem which involve the solution of Maxwell equations with boundary conditions. Numerical solutions of electrostatic problems within a non-conducting medium are based on the solutions of Poisson's equation 
\begin{equation}
\label{eq:Poisson}
{\mathbf \nabla} \cdot (\varepsilon {\mathbf \nabla} \Phi) = -Q
\end{equation}
where $\varepsilon$, $\Phi$ and  $Q$ are the electrical potential, permittivity of the medium and total charge in the considered region, respectively. However, if the medium has conductive regions, where no free charges and sources of charges are allowed, then, the solutions are given by
\begin{equation}
\label{eq:Laplace}
{\mathbf \nabla} \cdot (\sigma {\mathbf \nabla} \Phi) = 0
\end{equation}
where $\sigma$ is the conductivity of the region. When the medium is a mixture of these two cases, it consists of capacitive and resistive components. The solution is then given by a time dependent and complex electric potential in the region with the coupling of Eqs.\ (\ref{eq:Poisson}) and (\ref{eq:Laplace}).
\begin{align}
\begin{split}
{\mathbf \nabla} \cdot (\sigma {\mathbf \nabla} \Phi) + {\mathbf \nabla} \cdot (\frac{d}{dt} (\varepsilon {\mathbf \nabla} \Phi)) &= 0 \\
{\mathbf \nabla} \cdot ((\sigma + i w \varepsilon) {\mathbf \nabla} \Phi) &= 0
\label{eq:3b}
\end{split}
\end{align}
where  $i$ is $\sqrt{-1}$, no free charges are allowed in the region due to resistive component. Last two equations are the continuity and the Laplace equations for a medium with a complex dielectric constant.

Finite element method was used in order to solve the equations with the help of a field calculation software\cite{AceManual}. In this method, the region is meshed using an adaptive meshing technique with triangles and then the equations are solved for each triangle by interpolating the potential and its normal derivative (electric field) using boundary conditions\cite{Silvester1990}. In order to have a reliable numerical solution, linear, quadratic and cubic solution sets are compared for different mesh sizes and results of the cubic solutions are used when there were not any differences in the solutions. The boundary conditions were chosen due to symmetry of the unit cell, which were repeated in one direction $\sigma(\partial \Phi / \partial {\mathbf n}) = -j$, where $j$ is the current density, and rms voltage difference, $\Delta V$, of 1\ V were applied on the other boundaries, respectively. It is illustrated in Fig.\ \ref{unitcell}.

Frequency dependent capacitance was calculated using the total charges, $Q_t$ per unit length on one of the voltage applied boundaries by the Gauss' law,
\begin{equation}
\label{eq:C}
C(f) = \frac{Q_t(f)}{\Delta V} = \oint_l {\mathbf D} \cdot {\mathbf n}\ dl
\end{equation}
where,  ${\mathbf D}$ is the obtained electrical displacement normal to the line incriment $dl$ of the region, $l$ \cite{Jackson1975}, and it is equal to,
\begin{equation}
\label{eq:D}
{\mathbf D} = - \varepsilon \frac{\partial \Phi}{\partial {\mathbf n}} 
\end{equation}

Finally, frequency dependent real part of the relative permittivity, $\varepsilon'(f)$, was obtained by this value over the geometrical capacitance of the primitive cell, $C_0$ . 
\begin{equation}
\label{eq:eps}
\varepsilon'(f) = \frac{C(f)}{C_0}
\end{equation}

In the same way, frequency dependent conductance, $G(f)$ was calculated using the total normal current, $I_n$ per unit length on one of the voltage applied boundaries by Amp{\'e}re's law,
\begin{equation}
\label{eq:R}
G(f) = \frac{j_n(f)/l}{\Delta V} = \frac{ \oint_R {\mathbf I} \cdot {\mathbf n}\ dl}{l}
\end{equation}
where,  ${\mathbf I}$ is the obtained current to the line incriment $d{\mathbf l}$ of the region, $l$, and it is equal to,
\begin{equation}
\label{eq:I}
{\mathbf I} =  \sigma \frac{\partial \Phi}{\partial {\mathbf n}} 
\end{equation}
Frequency dependent imaginary part of the dielectric permittivity, $\varepsilon''(f)$ is then
\begin{equation}
\varepsilon''(f)=\frac{G(f)}{2 \pi \varepsilon_0 f}
\end{equation}


In the simulations, the material parameters of the media were $\varepsilon_1=4$ and $\varepsilon_2=40$, and conductivities $\sigma_1=10^{-12}~\Omega^{-1}$m$^{-1}$ and $\sigma_1=10^{-10}~\Omega^{-1}$m$^{-1}$. The boundary conditions were assigned such that when $\beta_{1,2}=\pi/2$ the inclusions were parallel to the electric field, $\mathbf{E}$. 
\begin{figure}[t]
      \psfragscanon
      \psfrag{LOG(OMEGA)}[c][c]{$\log(\omega)$}
      \psfrag{LOG(CHI)}[c][c]{$\log(\chi)$}
      \psfrag{0-0-a40-RE}[c][r]{$\textsf{I}$}
      \psfrag{0-0-a40-IM}[c][r]{$\textsf{II}$}
      \psfrag{0-30-a40-RE}[c][r]{$\textsf{III}$}
      \psfrag{0-30-a40-IM}[c][r]{$\textsf{IV}$}
      \psfrag{0-60-a40-RE}[c][r]{$\textsf{V}$}
      \psfrag{0-60-a40-IM}[c][r]{$\textsf{VI}$}
      \psfrag{0-90-a40-RE}[c][r]{$\textsf{VII}$}
      \psfrag{0-90-a40-IM}[c][r]{$\textsf{VIII}$}
      \includegraphics[width=6.5cm,angle=-90]{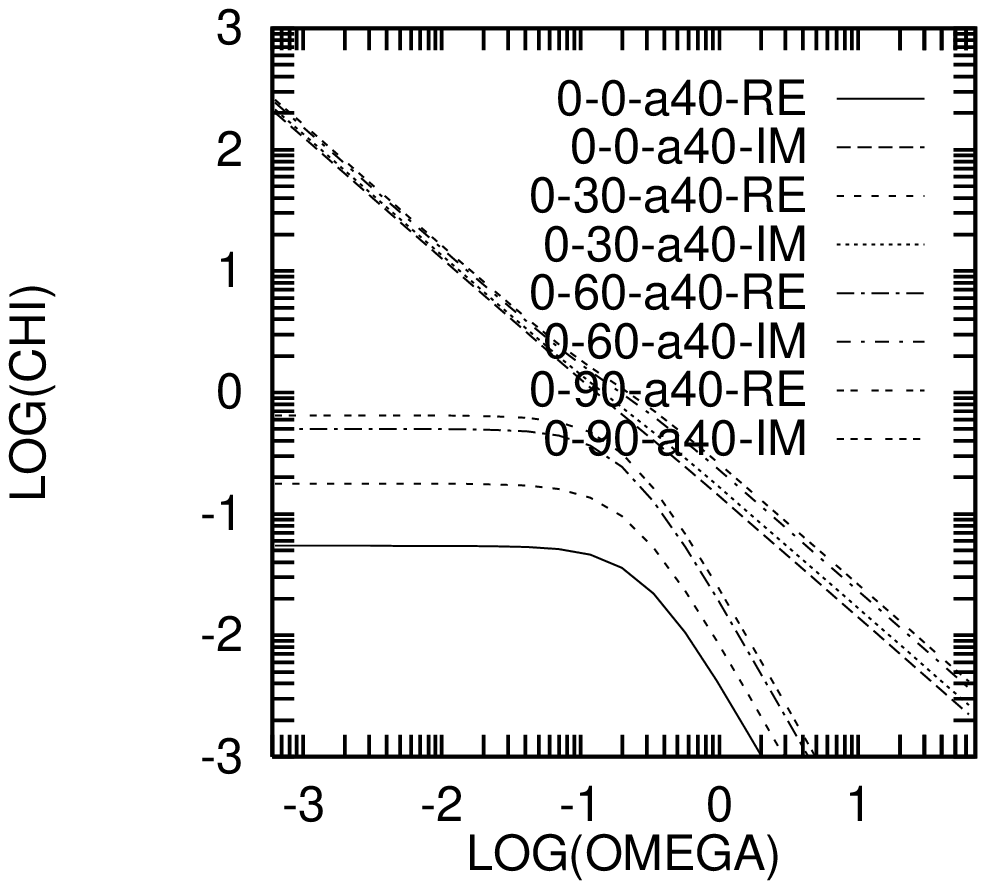}
      \psfragscanoff \\
      \psfragscanon
      \psfrag{90-0-a40-RE}[c][r]{$\textsf{I}$}
      \psfrag{90-0-a40-IM}[c][r]{$\textsf{II}$}
      \psfrag{90-30-a40-RE}[c][r]{$\textsf{III}$}
      \psfrag{90-30-a40-IM}[c][r]{$\textsf{IV}$}
      \psfrag{90-60-a40-RE}[c][r]{$\textsf{V}$}
      \psfrag{90-60-a40-IM}[c][r]{$\textsf{VI}$}
      \psfrag{90-90-a40-RE}[c][r]{$\textsf{VII}$}
      \psfrag{90-90-a40-IM}[c][r]{$\textsf{VIII}$}
      \psfrag{LOG(OMEGA)}[c][c]{$\log(\omega)$}
      \psfrag{LOG(CHI)}[c][c]{$\log(\chi)$}
      \includegraphics[width=6.5cm,angle=-90]{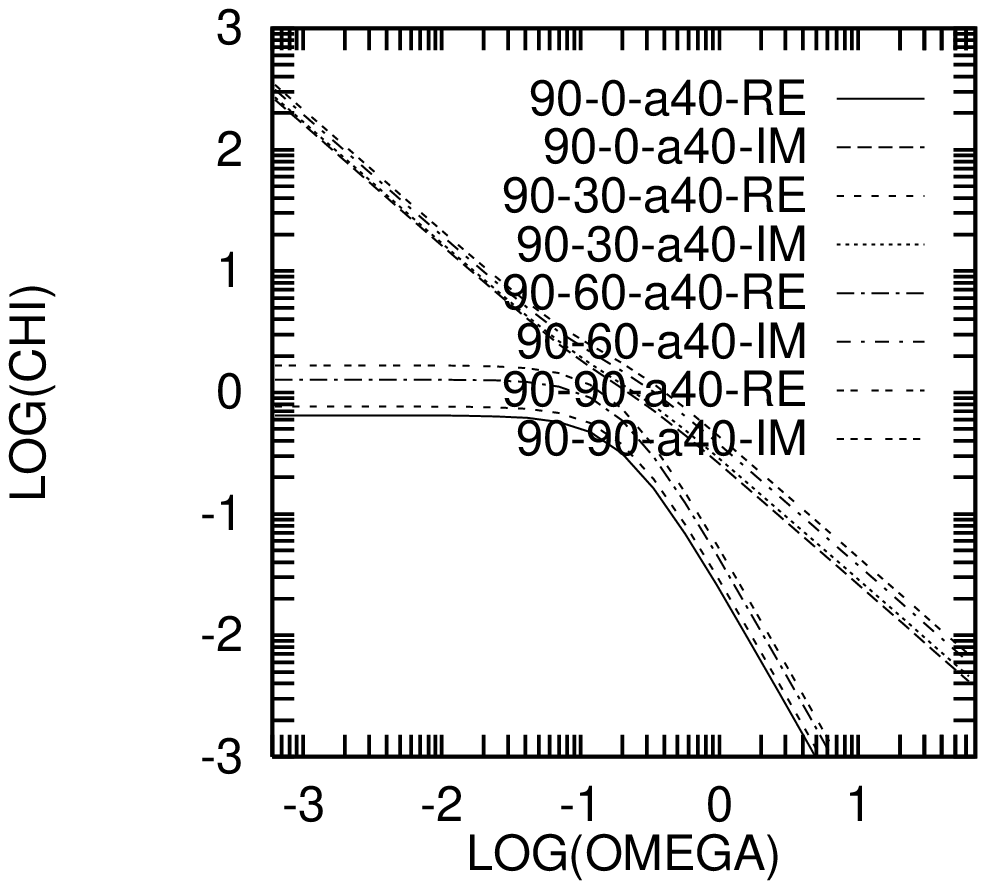}
      \psfragscanoff
    \caption{\label{fig:shape3}Complex dielectric susceptibilities, $\chi^\ast$, of a binary system with different orientations at $q=0.1$ when $e=0.980$. The orientation of one ellipse is kept constant (a) $\beta_1=0$ and (b) $\beta_1=\pi/2$, and the orientations of the nearest neighbors, $\beta_2$ are altered. The angle values are $\beta_2=[0,\pi/6,\pi/3,\pi/2]$ for the labels. The odd and even labels represent $\chi'$ and $\chi''$-values, respectively.}
\end{figure}

The influence of shape of inclusions were investigated by considering two $e$-values at $q=0.1$. Moreover, circular-inclusions, $e=0$, were also considered for comparison. In Fig.~\ref{fig:shape1}, the calculated complex dielectric susceptibilities, $\chi^\ast$, are presented for $e=0.980$ and $e=0.965$ while $\beta_1$ and $\beta_2$-values were kept constant, $\beta_{1,2}=\pi/2$ and $\beta_{1,2}=0$, respectively. When the $e$-value was close to 1, the shape of the inclusion was needle-like. Moreover, when the orientation of the inclusions were parallel to $\mathbf{E}$, $\beta_{1,2}=\pi/2$, they had not only higher $\chi'$-values, but also higher losses, $\chi''$, than the circular-shape. If the $\chi^\ast$-values of two elliptical-shapes were compared, the ellipse with higher $e$-value had the highest $\chi^\ast$-values. If the opposite case was taken into consideration, in which the orientations of the inclusions were perpendicular to $\mathbf{E}$, $\beta_{1,2}=0$, then, both shapes had similar $\chi'$-curves which were lower than that of the circular-shape. 
The $\chi''$-values of the all considered cases showed the ohmic losses which hid the losses due to interfacial polarization.  

\begin{figure}[t]
      \psfragscanon
       \psfrag{y}[c][c]{$\varepsilon_{hf}$}
      \psfrag{x}[c][c]{$\beta_2$}
      \psfrag{o}[c][c]{$0$}
      \psfrag{p}[][]{$\pi/2$}
      \psfrag{q}[c][c]{$\pi$}
      \includegraphics[width=6.5cm,angle=-90]{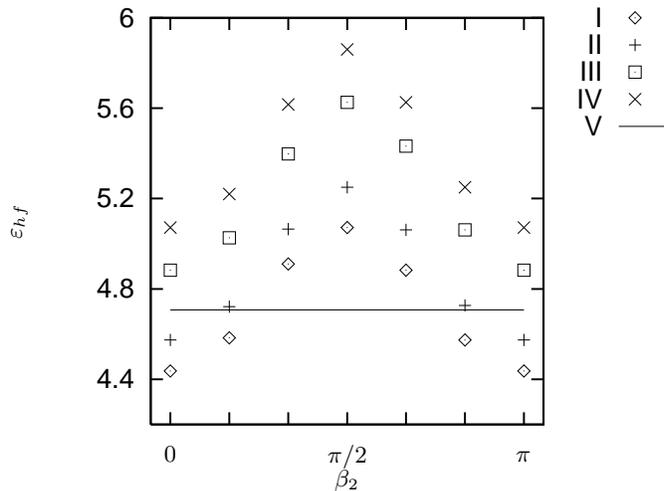}
      \psfragscanoff
    \caption{\label{fig:ellipse_eps_hf} Dielectric permittivity at high frequencies, $\varepsilon_{hf}$, of a binary system with different orientations of inclusions at $q=0.1$ when $e=0.980$. The $\beta_1$-value is kept constant and $\beta_2$-values are altered. The labels $\sf{I}$, $\sf{II}$, $\sf{III}$ and $\sf{IV}$ represent the cases when $\beta_1=[0,\pi/6,\pi/3,\pi/2]$, respectively. The solid-line ($\sf{V}$) is the $\varepsilon_{hf}$-value for a circular-inclusion.}
\end{figure}

In Fig.~\ref{fig:shape3}, a more general case, when $\beta_1\ne\beta_2$, is presented. When $\beta_1=0$, the influence of the orientational state, $\beta_2$, of the neighboring inclusions was affecting the $\chi^\ast$-values. The effect was, on the contrary, not that significant when $\beta_1=\pi/2$, as illustrated in Fig.~\ref{fig:shape3}. Finally, the influence of orientation of inclusions on dielectric permittivity at high frequencies, $\varepsilon_{hf}$ was considered, and is displayed in Fig.~\ref{fig:ellipse_eps_hf}. The $\varepsilon_{hf}$-values increased when the orientations of nearest neighbors were close to the direction of the applied field, $\mathbf{E}$.


In order to investigate the shape and orientation of inclusions, elliptical inclusions were considered. Needle-like inclusions had higher dielectric strengths, $\Delta\varepsilon$, and observable dielectric losses, $\chi''$, compared to the circular shapes. The effects were larger when the orientation of the longer-axis of the inclusions were parallel to the electric field. Moreover, when the orientations of the nearest neighbors were considered, higher dielectric strengths were observed for neighbors oriented close in the direction of the applied field.

\end{document}